\documentclass[twocolumn,showpacs,preprintnumbers,amsmath,amssymb]{revtex4}
\usepackage{epsfig}
\usepackage{graphics}
\usepackage{graphicx}% Include figure files
\usepackage{dcolumn}% Align table columns on decimal point
\usepackage{bm}% bold math

\begin{document}

\preprint{APS/123-QED}

\title{Bubbling route to strange nonchaotic attractor in a nonlinear series LCR 
circuit with a nonsinusoidal force}

% Force line breaks with \\
\author{D.~V.~Senthilkumar$^1$}
\author{K.~Srinivasan$^2$}
%\email{skumar@cnld.bdu.ac.in}
\author{K.~Thamilmaran$^1$}%
% \email{maran@cnld.bdu.ac.in}
 %\altaffiliation{skumar@cnld.bdu.ac.in}%Lines break automatically or can be forced with \\
\author{M.~Lakshmanan$^1$}%
\email{lakshman@cnld.bdu.ac.in}
\affiliation{%
$^1$Centre for Nonlinear Dynamics, Department of Physics,
Bharathidasan University, Tiruchirapalli - 620 024, India \\
 }%
\affiliation{%
$^2$Department of Physics,
 National Institute of Technology,
Tiruchirappalli - 620 015, India\\
}%

\date{\today}% It is always \today, today,

\begin{abstract} 

We identify a novel route to the birth of a strange nonchaotic attractor
(SNA) in a quasiperiodically forced electronic circuit with a nonsinusoidal
(square wave) force as one of the quasiperiodic forces through
numerical and experimental studies. We find  that bubbles appear in the
strands of the quasiperiodic attractor due to the instability induced by the
additional square wave type force.  The bubbles then enlarge and get
increasingly wrinkled as a function of the control parameter.  Finally, the
bubbles get extremely wrinkled (while the remaining parts of the strands of the
torus remain largely unaffected) resulting in the birth of the 
SNA which we term as the \emph{bubbling route to SNA}. We characterize and confirm this birth 
from both experimental and numerical data by maximal
Lyapunov exponents and their variance, Poincar\'e maps, Fourier amplitude
spectra and spectral distribution function. We also strongly confirm
the birth of SNA via the bubbling
route by the distribution of the finite-time Lyapunov
exponents. 

\end{abstract}

\pacs{05.45.-a,05.45.Pq,05.45.Ac,05.45.Df,95.10.Fh}% PACS, the Physics and Astronomy
\maketitle

\section{\label{sec:level1}Introduction}

Strange nonchaotic attractors (SNAs) are considered as typical structures of
quasiperiodically forced nonlinear systems. They are geometrically strange (that
is they are fractal in nature)  just like the chaotic attractors, while all
their Lyapunov exponents are either zero or negative which ensure that the
underlying dynamics is nonchaotic. Further, due to their fractal nature, the SNAs
are characterized by aperiodic oscillations. Following the pioneering work of
Grebogi et al.~\cite{cgeo1984}, SNAs have been 
extensively  investigated theoretically in several dynamical systems 
~\cite{fjreo1987,abeo1985,mdcg1989,jfhwld1991,tyycl1996, tkjw1993,avml2000,
avml1997,apvm1997,aspuf1995,vsatev1996,
tnkk1996,avml2001,brheo2001,jfhsmh1994,aprr1999,csztlc1997,sykwl2004,tkloc1997}.
The existence of SNAs has also been demonstrated
experimentally~\cite{tykb1997,wldmls1990,avkm1999,ktdvsk2006} in a few
physically relevant situations. As a consequence, several routes (scenarios
having distinct signatures) to SNAs have been reported theoretically. These
include Heagy-Hammel route~\cite{jfhsmh1994}, gradual fractalization
route~\cite{tnkk1996}, various types of intermittency
routes~\cite{avkm1999,apvm1997,brheo2001,sykwl2004}, blowout  bifurcation
route~\cite{tyycl1996}, etc. As all these bifurcation scenarios (routes to SNAs)
have been well established in the literature, we summarize the different scenarios for
the formation of SNAs along with their distinct signatures/mechanisms in
Table~I. Reviews on SNAs can  be found in
Refs.~\cite{tkjw1993,apssn2001,ufsk2006}.

As mentioned above, while extensive numerical studies on the birth of
SNAs via different routes are available in the literature
~\cite{fjreo1987,abeo1985,mdcg1989,jfhwld1991,tyycl1996, tkjw1993,avml2000,
avml1997,apvm1997,aspuf1995,vsatev1996,
tnkk1996,avml2001,brheo2001,jfhsmh1994,aprr1999,csztlc1997,sykwl2004,tkloc1997},
only a few experimental realizations of them exist
~\cite{tykb1997,wldmls1990,avkm1999,ktdvsk2006}. In particular, these exotic
attractors were confirmed by an experiment consisting of a quasiperiodically
forced, buckled, magneto-elastic ribbon~\cite{wldmls1990}. SNAs were also
realized in analog simulations of a multistable potential~\cite{tzfm1992}, and
in a neon glow discharge experiment~\cite{wxdhd1997}. These attractors were also
shown to be related to the Anderson localization phenomenon in the Schr$\ddot
{\text o}$dinger equation with a quasiperiodic
potential~\cite{jakis1997,aprr1999}. Very recently SNAs have also been observed in an excitable chemical system, namely a three electrode electrochemical cell~\cite{ruiz2007}.  In this  connection, from an experimental
point of view, nonlinear electronic circuits with suitable quasiperiodic forces
turn out to be especially useful dynamical systems for the identification and
study of SNAs. For example, Type-I intermittency route to SNA was reported
in a quasiperiodically forced Murali-Lakshmanan-Chua circuit~\cite{avkm1999}.
Recently, three prominent routes, namely Heagy-Hammel, fractalization and
type-III intermittency routes to SNAs, have been identified and reported in a
quasiperiodically forced negative conductance series LCR circuit with a
diode~\cite{ktdvsk2006} both experimentally and numerically by some of the
present authors.

In almost all the above studies, as a general rule, the driving forces are 
assumed to be sinusoidal in nature.  Naturally the question arises as to  what
happens to the dynamics when one or both of the driving forces are nonsinusoidal
but periodic.  Can new routes to the birth of SNAs emerge in such a scenario? 
In order to answer these questions, we consider the quasiperiodically driven
negative conductance series LCR circuit with a diode (which was investigated in 
~\cite{ktdvsk2006}) and unravel the dynamics of the circuit with  one of the
forces taken as a square wave force (nonsinusoidal) for suitable parameter
values.  The main reason for choosing square wave as one of the driving forces
is  its bistable nature.  Bistability is responsible for hysteresis in many
physical and technical systems~\cite{sgps2004,ddrl2006,arbej1991,vcmm2000}.  Further, the square wave has also been used for inducing chaos
in certain dynamical systems~\cite{zmgwyl2004}. For example a 10
MHz square wave optical message was injected into a ring laser to produce
high-dimensional chaotic light~\cite{gdvrr1998}.  Thus the study of the present circuit has considerable relevance in understanding SNA transitions. 

\begin{table}
\caption{Routes and mechanisms for the formation of SNAs}
%\small
%\begin{tabular}{|p{7cm}|p{7cm}|}
%\hline
%{\bf \emph \;\;\;\;\;\;\;\;\;\;\;\;\;\;Type of route} & {\bf \emph \;\;\;\;\;\;\;\;\;\;\;\;\;\;\;\;Mechanism}  \\
\begin{ruledtabular}
\begin{tabular}{p{4cm}p{10cm}}
{\bf \emph \;\;\;\;\;\;\;Type of route} & {\bf \emph \;\;\;\;\;\;\;\;\;\;Mechanism}  \\
\hline
Heagy-Hammel \cite{jfhsmh1994}  & Collision of period-doubled torus \\
&with its unstable parent\\
%\hline  
Gradual Fractilization \cite{tnkk1996} & Increased wrinkling of torus  \\
& without any interaction with \\
& nearby periodic orbits\\
%\hline
%On-off intermittency \cite{tyycl1996} &Loss of transverse stability of \\
%&torus\\
%\hline
Type-I intermittency \cite{apvm1997} & Due to saddle-node bifurcation, \\
& a torus is replaced by SNA\\
%\hline
Type-III intermittency \cite{avkm1999}  & Subharmonic instability\\
&\\
%\hline
Crisis-induced intermittency & Doubling of destroyed torus\\
 \cite{avml2001} &involves a kind of sudden\\
& widening of the attractor\\
%\hline
Homoclinic collision \cite{aprr1999} & Homoclinic collisions of the \\
& quasiperiodic orbits\\
%\hline
Blowout bifurcation \cite{tyycl1996} & Due to changes in sign of the \\
& Lyapunov exponent $\Lambda_T$ transverse \\
& to the invariant subspace $S$\\
%\hline.
Quasiperiodic route \cite{aspuf1995,avml1997} & Collision between a stable \\
& and unstable torus\\
%\hline
%\end{tabular}
\end{tabular}
\end{ruledtabular}
\end{table} 

In the present proposed circuit with a square wave force as one of the
quasiperiodic forces in addition to a sinusoidal force, we have identified a new
route for the formation of SNA which we term as the {\bf bubbling
route} to SNA. \emph{In this route bubbles appear in the strands of the torus as
a function of the control parameter, then the sizes of the bubbles increase with
the value of the control parameter and subsequently strands of the bubbles  are
increasingly wrinkled resulting in the birth of SNA (while the remaining parts of
the strands of the torus outside the bubbles remain largely unaffected). The
mechanism for this route is that the quasiperiodic orbit becomes increasingly
unstable in its transverse direction as a function of the control parameter
which is induced by the square wave type quasiperiodic force resulting in an
increase in the size of the doubled strands (bubbles) in certain parts of the
main strand and then the doubled strands become extremely wrinkled (without a
complete doubling of the entire  main strand) resulting in the SNA}. In addition
to this we have also observed four other prominent routes in the same
circuit which include fractalization, fractalization followed by intermittency,
intermittency and Heagy-Hammel routes, the details of which will be published
elsewhere. 

In order to confirm the existence of the bubbling route to SNA in the proposed
circuit, we first
present a detailed  numerical analysis of the dynamical equations of the circuit
in a rescaled form for suitable values of the parameters using various
qualitative and quantitative measures to establish this route.  These include
Poincar\'e surface of section, Fourier spectrum, largest Lyapunov exponent and
its variance, spectral distribution function and distribution of finite time
Lyapunov exponents. A short account of these measures is given in
Appendix~\ref{a1}. Next, we confirm the results experimentally by the  phase
portraits of the quasiperiodic attractors and SNAs for the corresponding values
of the circuit parameters, again with appropriate quantification measures, to
establish the existence of torus and the birth of SNA through the bubbling
route. 

The paper is organized as follows.  In Sec.~II, we discuss the circuit
realization of the quasiperiodically forced negative conductance series LCR
circuit with diode using a  sinusoidal and a nonsinusoidal (square wave) forcing
as quasiperiodic forces. In Sec.~III, we describe the phase diagram of the
circuit where the  regions corresponding to the different dynamical transitions
to SNAs are delineated as a function of the control parameters, based on our
numerical  analysis. The birth of SNA via the bubbling route as confirmed in the
numerical analysis is discussed in Sec.~IV.   Experimental confirmation of the
bubbling route to SNA is presented in Sec.~V. Finally, we summarize our results in
Sec.~VI. Appendix~\ref{a1} contains a short summary on the identification and
characterization of SNA and the associated routes.

\section{\label{cr}Circuit realization}

In this section details about the proposed circuit are presented and the circuit
equations are written in terms of the circuit variables. Then the circuit
equations are transformed into dimensionless equations (normalized equations)
using appropriate rescaled variables for a convenient numerical analysis.

\subsection{Experimental realization: Circuit Equations}
\begin{figure}
\centering
\includegraphics[width=1.0\columnwidth]{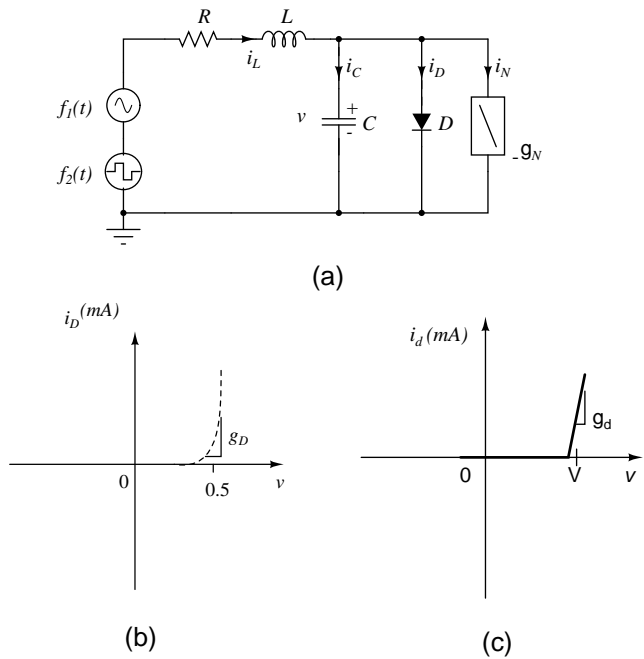}
\caption{\label{fig1}(a) Circuit realization of a simple nonautonomous circuit. 
Here $D$ is the {\it{p-n junction}} diode and $g_N$ is negative conductance.
The external emfs are $f_1(t)=E_{f1} \sin(\omega_{f1}t)$  and
$f_2(t)=E_{f2}sgn(\sin(\omega_{f2}t))$. The values of the circuit elements are
fixed as $L=50.3~mH$, $C=10.35~nF$, $R=1900ohms$, $E_{f2}=400mV$ and
$\omega_{f2}=17033Hz$.    The forcing amplitude $E_{f1}$ and its frequency
$\omega_{f1}$ are chosen as the control parameters. 
(b) $i-v$ characteristics of the {\it{p-n junction}} diode and (c) two
segment piecewise-linear function.}
\end{figure}

We consider the simple second-order nonlinear dissipative nonautonomous negative
conductance series LCR circuit with a sinusoidal voltage generator, $f_1(t)$,
introduced by us very recently~\cite{ktdvsk2005,ktdvsk2006} along with a second
nonsinusoidal force, $f_2(t)$, as shown in Fig.~\ref{fig1}a.

The circuit consists of a series LCR network, forced by a sinusoidal voltage
generator, $f_1(t)$, and a nonsinusoidal (square wave) voltage generator, $f_2(t)$ (HP 33120A
series).  Two extra components, namely a {\it{p-n junction}} diode
(D) and a linear negative conductor $g_N$, are connected in parallel to the
forced series LCR circuit.  The negative conductor used here is a standard
op-amp based negative impedance converter (NIC).  The diode operates as a
nonlinear conductance, limiting the amplitude of the oscillator.  In
Fig.~\ref{fig1}a, $v,i_L$ and $i_D$ denote the voltage across the capacitor $C$,
the current through the inductor $L$ and the current through the diode $D$,
respectively.  The actual $v-i$ characteristic of the diode (Fig.~\ref{fig1}b)
is approximated  by the usual two segment piecewise-linear function  (
Fig.~\ref{fig1}c) which facilitates numerical analysis considerably.   The state
equations governing the presently proposed circuit (Fig.~\ref{fig1})  are a set
of two first-order nonautonomous differential equations
\begin{subequations} 
\begin{align} 
C\frac{dv}{dt} =&\,i_L- i_D +g_{N}v, \\
L\frac{di_L}{dt} =&\,-Ri_L-v + E_{f1} \sin( \omega_{f1} t )\nonumber\\
&\,+E_{f2}sgn(\sin(\omega_{f2}t)),
\end{align} 
where
\begin{eqnarray}
i_D(v)=
\left\{
\begin{array}{cc}
g_D(v-V),& v \ge V, \\
0,& v < V. \\
\end{array} \right.
\end{eqnarray}
\label{nor_eq}
\end{subequations} 
Here $g_D$ is the slope of the characteristic curve of the diode, $E_{f1}$ and
$E_{f2}$ are the amplitudes, $\omega_{f1}$ and $\omega_{f2}$ are the angular
frequencies of the forcing functions $f_1(t)=E_{f1} \sin(\omega_{f1}t)$ and 
$f_2(t)=E_{f2}sgn(\sin(\omega_{f2}t))$, respectively.  In the absence of
$f_2(t)$, the circuit (Fig.~\ref{fig1}a) has been shown to exhibit chaos and
also strong chaos not only through the familiar period-doubling route but also
via torus breakdown followed by period-doubling bifurcations~\cite{ktdvsk2005}.
Here our aim is to investigate the effect of the second square wave type
external forcing on the dynamics and to identify different types of transitions
to SNAs. 

In order to select the appropriate set of  experimental parameters for which
SNAs can be actually observed, we first carry out a detailed numerical
simulation (as pointed out below) which then serves as a guide and a
characterizer for experimental investigation.  Using such an analysis, the
values of the diode conductance $g_D$, negative conductance $g_N$ and  break
voltage $V$ are fixed as $1313\mu S$, $-0.45mS$ and $0.5V$,
respectively. We have fixed the actual experimental values of the resistance $R$, 
inductance
$L$, capacitance $C$, external frequency $\omega_{f2}$ and forcing strength
$E_{f2}$ of the square wave as $1900 ohms$, $50.3mH$, $10.35nF$, $17033Hz$ and $400mV$,
respectively, while we vary the amplitude $E_{f1}$ and the frequency
$\omega_{f1}$  of the sinusoidal force as control parameters in order to observe
the various dynamical states. The forcing functions $f_1(t)$ and $f_2(t)$ are
obtained from two separate function generators of the type $HP33120A$.

\subsection{Numerical analysis: Normalized equations}
In order to study the dynamics of the circuit in detail, Eq.~(\ref{nor_eq}) can
be converted into a convenient normalized form for numerical analysis by using
the following rescaled variables and parameters, $\tau=t/{\sqrt{LC}}$,
~$x=v/V$, ~$y=(i_L/V)(\sqrt{L/C})$, ~$E_1=E_{f1}/V$, $E_2=E_{f2}/V$,
~$\omega_1=\omega_{f1}\sqrt{LC}$, ~$\omega_2=\omega_{f2}\sqrt{L C}$,
~$a=R\sqrt{C/L}$, ~$b=g_N\sqrt{L/C}$, and ~$c=g_D\sqrt{L/C}$.\\ The normalized
evolution equation so obtained from Eq.~(\ref{nor_eq}) is
\begin{subequations}
\begin{eqnarray}
\dot{x} & = & y + f(x),   \\ 
\dot{y} & = & -x -ay + E_1 \sin(\phi)+E_2sgn(\sin(\theta)), \\
\dot{\phi} & = & \omega_{1}, \\
\dot{\theta} & = & \omega_{2}, 
\end{eqnarray}
where
\begin{eqnarray}
f(x)=
\left\{
\begin{array}{cc}
(b-c)x+c,& x \ge 1, \\
bx,& x < 1. \\
\end{array} \right.
\end{eqnarray}
\label{cir_eq}
\end{subequations}
Here dot stands for differentiation with respect to $\tau$.
Eq.~(\ref{cir_eq}) is then numerically integrated using Runge-Kutta fourth order
routine to identify the different dynamical scenarios corresponding to 
different values of the rescaled parameters. Various interesting dynamical
transitions of the Eq.~(\ref{cir_eq}) are  described below.

\begin{figure}
\centering
\includegraphics[width=1.0\columnwidth]{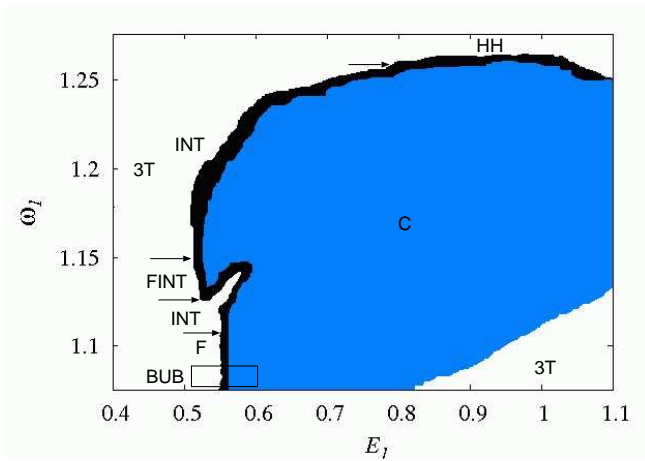}
\caption{\label{fig2}(Color online) Numerical phase diagram in the
$(E_{1}-\omega_{1})$ plane for the circuit given in Fig.~\ref{fig1}, represented
by Eqs.~(\ref{cir_eq}). {\bf 3T} correspond to period-3 torus, {\bf F}, {\bf
BUB},  {\bf FINT}, {\bf INT}, and {\bf HH} denote the birth of SNAs through
gradual fractalization, bubbling, fractalization followed by intermittency,
intermittency and Heagy-Hammel routes, respectively.  {\bf C} corresponds to the
chaotic attractor. Arrows indicate the transition regions between  two different
types of routes to SNA.}
\end{figure}

\section{\label{sectp} Two parameter Phase diagram}

The parameter space of the amplitude of the external forcing $E_1$ and the
frequency $\omega_1$ of the sinusoidal forcing is scanned first numerically in
the range of $E_1\in(0.4,1.1)$ and $\omega_{1}\in{(1.075,1.275)}$ to pinpoint
different dynamical behaviors and more specifically for the occurrence of
SNAs through different routes.  From this analysis, various dynamical
transitions are determined as a function of the amplitude of the external
forcing $E_1$ and its frequency $\omega_1$. Further, these dynamical behaviors
and their transitions are also confirmed experimentally for the corresponding
values of the experimental parameters of the circuit given in Fig.~\ref{fig1}.

\subsection{Numerical analysis}

To start with, we first demarcate the parameter space  $(E_{1},\omega_{1})$, by
numerically integrating Eqs.~(\ref{cir_eq}), into quasiperiodic, strange
nonchaotic and chaotic regimes by using the various qualitative and quantitative
measures  as discussed in the Appendix~\ref{a1}. The numerical phase diagram is
shown in Fig.~\ref{fig2} for $E_1\in(0.4,1.1)$ and
$\omega_{1}\in{(1.075,1.275)}$. The various dynamical behaviors indicated in the
phase diagram (Fig.~\ref{fig2}) and the interesting dynamical transitions are
elucidated in the following.

\begin{figure}
\centering
\includegraphics[width=1.0\columnwidth]{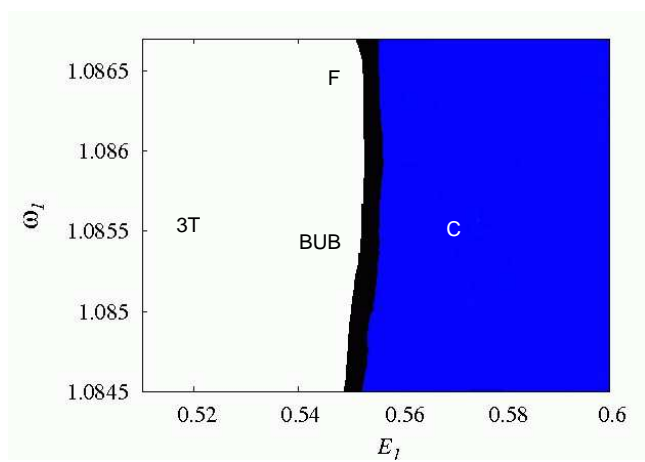}
\caption{\label{fig2a}(Color online) Blow up picture of Fig.~\ref{fig2} in the
bubbling transition regime indicated as {\bf BUB}.}
\end{figure}

Transitions from quasiperiodic attractor to SNA and subsequently to chaotic
attractor occur on increasing the value of the amplitude of
the sinusoidal force $E_{1}$ for fixed value of its frequency $\omega_{1}$.
Strange nonchaotic attractors created through different mechanisms are found to
occur for different values of the frequency $\omega_{1}$ of the sinusoidal
force. Now, we will outline the ranges of values of the frequency $\omega_{1}$
for which  SNAs arise from quasiperiodic attractors through different mechanisms
on increasing the value of the amplitude $E_{1}$ of the sinusoidal force.

Strange nonchaotic attractor created through the newly proposed route, namely
the bubbling route, is identified in the range of frequency $\omega_{1}
\in(1.085,1.086)$.  Here bubbles appear in the strands of period-3 torus and
then the bubbles get increasingly wrinkled  in the range of the amplitude
$E_{1}\in(0.54,0.55)$ of the sinusiodal forcing resulting in SNA. This
phenomenon is named as the bubbling transition to SNA and it is denoted as {\bf BUB}
in  Fig.~\ref{fig2}.  A blow up of the two parameter space corresponding to the
bubbling transition is shown in Fig.~\ref{fig2a}. Further increase in the value
of $E_{1}$ ends up in the chaotic behavior indicated as {\bf C} in 
Figs.~\ref{fig2} and ~\ref{fig2a}. Strange nonchaotic attractor created through
gradual fractalization ({\bf F}) of period-3 (3T)  quasiperiodic attractor is
identified for $\omega_{1} \in(1.086,1.111)$ as a function of the
amplitude  $E_{1} \in (0.5,0.55)$. Intermittency route ({\bf INT}) is found to
be exhibited in the range of frequency $\omega_{1} \in(1.111,1.1268)$ on
increasing $E_1$ in the range $E_{1} \in (0.5,0.55)$ and also for 
$\omega_{1} \in (1.1512,1.2615)$ when $E_{1} \in (0.5,0.8)$. When the frequency 
$\omega_{1} \in(1.1268,1.1512)$, gradual fractalization
is followed by intermittency phenomenon on increasing the value of the amplitude of the external
forcing $E_{1}$. It is marked as ({\bf FINT}) in  Fig.~\ref{fig2}. Torus
doubling bifurcation from a period-3 torus (3T) to a period-6 (6T) torus and
then to SNA via the Heagy-Hammel ({\bf HH}) mechanism is found to occur in the
range of  $\omega_{1} \in(1.2501,1.2615)$ on decreasing $E_{1}$ in the range
$E_{1}\in(1.1,0.8)$. The transition regions between the above mentioned
dynamical regimes are indicated by arrows in Fig.~\ref{fig2} which are
fixed by scanning the frequency $\omega_1$ of the sinusoidal force at its fourth
decimal place. However, we do not draw a distinct boundary between any two
scenarios because it requires a much detailed numerical analysis on a finer
parameter scale. 

\subsection{Experimental investigation}

It has additionally been confirmed that the above dynamical behaviors are also exhibited by
the experimental circuit for the corresponding values of the circuit parameters
$E_{f1}(=V\cdot E_1)$ and $\omega_{f1}(=\sqrt{(C/L)}\cdot\omega_1)$ by examining
the two dimensional projections of the corresponding attractors obtained by
measuring the voltage $v$ across the capacitor $C$ and the current $i_L$ through
the inductor $L$ which are connected to the $X$ and $Y$ channels of an
oscilloscope, respectively. Here $V$ is the break voltage.  Then, a live picture
of the corresponding power spectrum obtained from a digital storage oscilloscope
(HP 54600 series) of the projected attractor has also been used to distinguish
the different attractors. In addition to this, the experimental data of the
corresponding attractors recorded using a 16-bit data acquisition system
[AD12-16U(PCI)EH] at the sampling rate of 200 kHz have been analyzed
quantitatively using the different quantification measures, namely the spectral
distribution function and the distribution of finite time Lyapunov exponents. 
This information  is then utilized (i) to pinpoint the different dynamical
behaviors, (ii) to distinguish the SNAs created through different mechanisms and
(iii) also to compare them with the results of numerical simulation.  In the
following, we will describe only the novel bubbling transition  in detail by
both numerical simulation  and experimental realization, while the results of
other known routes will be published elsewhere.

\section{Bubbling route to SNA: Numerical Analysis}

As noted above, in this new route, the bubbles appear in  the strands of the torus as the value
of the amplitude $E_{1}$ of the sinusoidal forcing is increased for a fixed
value of its frequency $\omega_{1}$. The sizes of the bubbles increase further
on increasing the amplitude $E_{1}$  and the bubbles  increasingly get wrinkled
(while the remaining parts of the strands of the torus outside the bubbles
remain largely unaffected) resulting in the birth of SNA. This  bubbling route 
is observed in the rather narrow range of frequency $\omega_{1} \in
(1.085,1.086)$ as a function of the amplitude of the sinusoidal forcing 
$E_{1} \in (0.5,0.55)$ indicated as {\bf BUB} in Figs.~\ref{fig2}
and ~\ref{fig2a}.  It is to be noted that this route is significantly different
from the well known fractalization route~\cite{tnkk1996}, where the entire
strands of the n-period torus will continuously deform and  get extremely
wrinkled  as a function of the control parameter. The formation of SNA through
this novel bubbling route has been identified in the literature for the first
time to the best of our knowledge. We have used both qualitative and 
quantitative measures, which are indicated in the Appendix \ref{a1},
to confirm the new route. The qualitative proof is given
through the Poincar\'e surface of section by distinguishing geometrically
between quasiperiodic attractors and SNAs. The quantitative confirmation is provided using three different measures:  (i) The largest Lyapunov exponents and its variance are used to distinguish between
torus and SNA, and SNA and chaotic attractors. (ii) Scaling laws deduced from the distribution function
for quasiperiodic attractors and SNAs are used to distinguish them. (iii) Finally, different routes to SNAs are also
distinguished by the different distributions of local Lyapunov exponents. More
information on the characterization is given in the Appendix \ref{a1}. In the following we provide details of
the confirmation of the bubbling route.

\begin{figure}
\centering
\includegraphics[width=1.0\columnwidth]{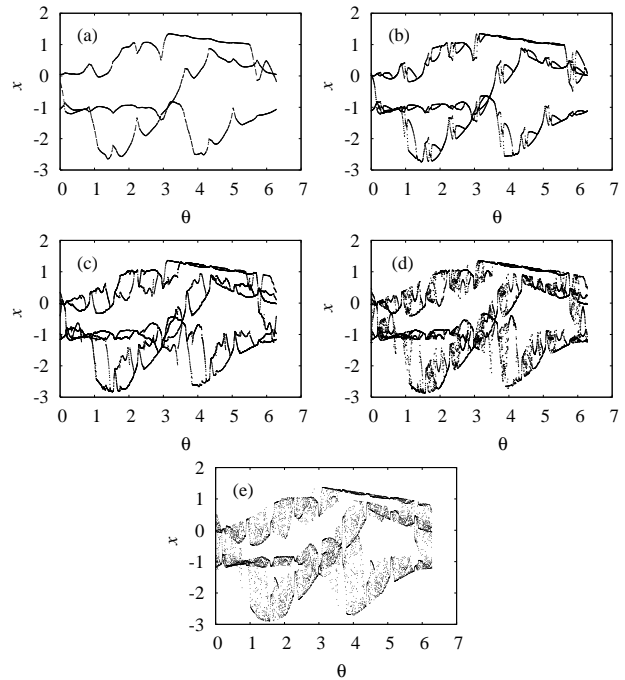}
\caption{\label{fig10} Projection of the numerically simulated Poincar\'e
surface of section  of the attractors of  Eqs.~(\ref{cir_eq}) in the $(\phi,
x)$ plane for a fixed value of the frequency of the sinusoidal forcing,
$\omega_{1}=1.0852$, as a function of its amplitude $E_{1}$ indicating the
transition from quasiperiodic attractor to SNA through bubbling route: (a)
period-3 torus (3T) for $E_{1}=0.5$, (b) bubbled strands of period-3 torus
(3T) for $E_{1}=0.52$, (c) enlarged bubbles in the strands of period-3 torus
(3T) for $E_{1}=0.54$, (d) fractalized bubbles  for $E_{1}=0.546$ with
the remaining parts (away from bubble) of the strands  unaffected and (e) chaotic attractor (widely interspersed bubbles) for $E_{1}=0.56$.}
\end{figure}

\begin{figure}
\centering
\includegraphics[width=1.0\columnwidth]{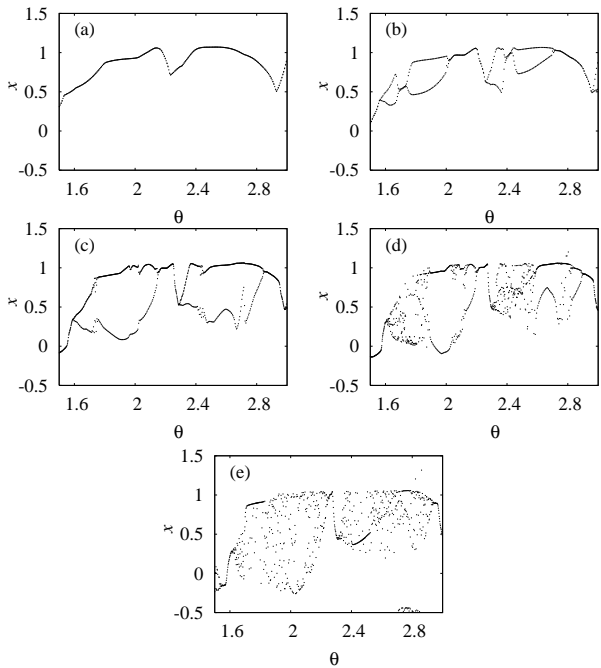}
\caption{\label{fig10a} Enlarged figures of Figs.~\ref{fig10} to show the
bubbling transition to strange nonchaotic attractor.}
\end{figure}

\subsection{Poincar\'e surface of section plots and power spectra}

\begin{figure}
\centering
\includegraphics[width=1.0\columnwidth]{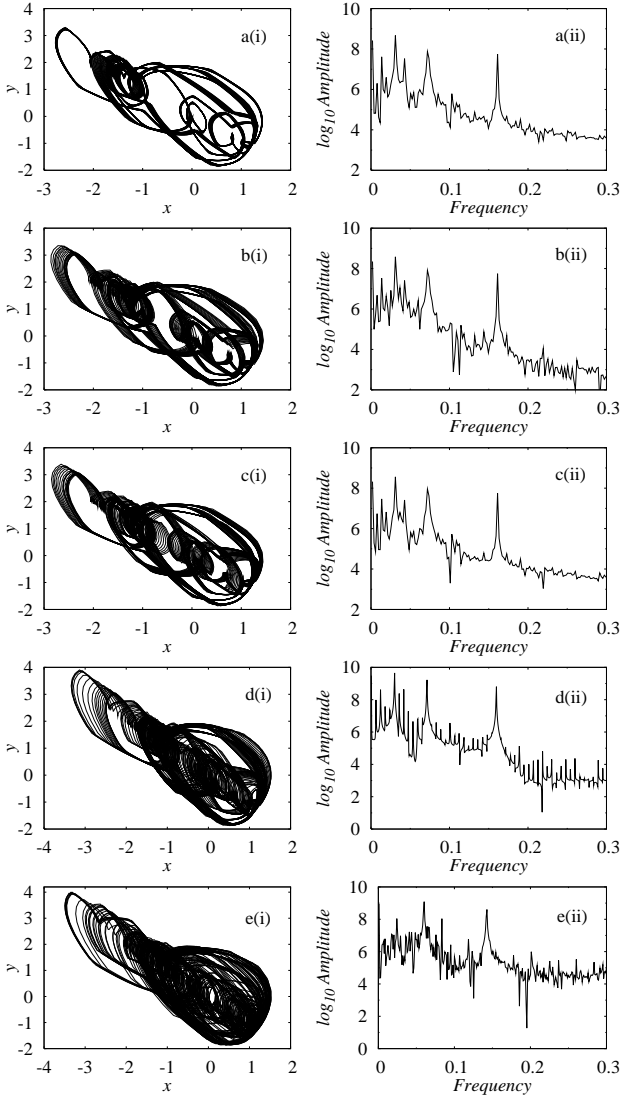}
\caption{\label{fig11} Projection of the numerically simulated attractors and
their power spectrum of  Eqs.~(\ref{cir_eq}) for the same values of  the
frequency $\omega_{1}$ and the amplitude $E_{1}$ of the sinusoidal forcing as
in Figs.~\ref{fig10}. (a) period-3 torus (3T), (b) bubbled period-3 torus,
(c) period-3 torus with enlarged bubbles, (d) fractalized bubbles  (SNA) and (e) chaotic attractor : (i)
phase portrait in the $(x,y)$ space; (ii) power spectrum.}
\end{figure}

We have fixed the value of the frequency of the sinusoidal forcing as
$\omega_{1}=1.0852$ for illustration and varied its amplitude in the range
$E_{1}\in(0.5,0.55)$ to elucidate the emergence of bubbling route to SNA in the
present system (\ref{cir_eq}).  The Poincar\'e surface of section plot of the three
strands corresponding to period-3 torus for the value of $E_{1}=0.5$ is shown in
Figs.~\ref{fig10}a and \ref{fig10a}a. The corresponding phase portrait and power
spectrum are depicted in Figs.~\ref{fig11}a(i) and~\ref{fig11}a(ii),
respectively. As the value of the amplitude $E_{1}$ is increased further,
bubbles start to appear in all the three strands  starting from $E_{1}=0.516$.
These are shown in Figs.~\ref{fig10}b and \ref{fig10a}b for $E_{1}=0.52$ and the
corresponding phase portrait and power spectrum are shown in
Figs.~\ref{fig11}b(i) and~\ref{fig11}b(ii), respectively. Further increase in
the value of $E_{1}$ results in an increase in the size of the bubbles as shown
in Figs.~\ref{fig10}c and \ref{fig10a}c for the value of $E_{1}=0.54$, whose
phase portrait and power spectrum are shown in Figs.~\ref{fig11}c(i)
and~\ref{fig11}c(ii), respectively. Beyond the value of $E_{1}=0.54$, the
strands of bubbles deform and get increasingly wrinkled (while the other parts
of the strands  of period-3 torus outside the bubbles remain unaltered as seen
in Fig.~\ref{fig10a}d) leading to the formation of SNA as depicted in
Fig.~\ref{fig10}d for the value of $E_{1}=0.546$.  The phase portrait and power
spectrum for this value of $E_{1}$ are shown in  Figs.~\ref{fig11}d(i)
and~\ref{fig11}d(ii), respectively.  Finally, to confirm that the SNA transits to a chaotic attractor beyond $E_{1}=0.55$, we have depicted the Poincar\'e surface of  section of the latter in Figs.~\ref{fig10}e and \ref{fig10a}e with the corresponding attractor and power spectrum in Figs.~\ref{fig11}e for $E_{1}=0.56$.  

The mechanism for the bubbling route is that the quasiperiodic orbit becomes
increasingly unstable in its transverse direction as a function of the control
parameter $(E_1)$, resulting in the formation of the doubled strands (bubbles),
as seen in Figs.~\ref{fig10}b and \ref{fig10a}b, in certain parts of the main
strands. This instability of the quasiperiodic attractor arises due to the
presence of the square wave pulse (finite amplitude for finite durations).
Further increase in the value of the amplitude of the forcing $(E_1)$ results
in  an increase in the size of the doubled strands (bubbles) as shown in
Figs.~\ref{fig10}c and \ref{fig10a}c, and then the doubled strands become
extremely wrinkled (without a complete doubling of the entire  main strand)
resulting in the SNA as depicted in  Figs.~\ref{fig10}d and \ref{fig10a}d. 

We now provide quantitative confirmation of the above results to distinguish between torus and SNA, and SNA and chaos.

\subsection{Largest Lyapunov exponent and its variance}
\begin{figure}
\centering
\includegraphics[width=1.0\columnwidth]{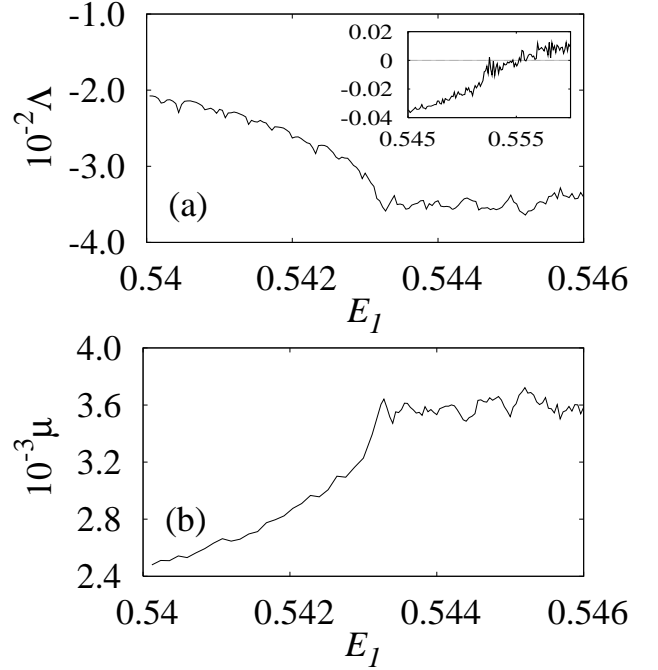}
\caption{\label{fig12} Transition from torus to SNA through bubbling
route for the same value of frequency as in Figs.~\ref{fig10} 
and in the range of amplitude $E_{1}\in(0.54,0.546)$ obtained numerically. 
(a) Largest Lyapunov exponent ($\Lambda$) and (b) its variance ($\mu$).  Inset in (a) depicts transition from SNA to chaos for $E_{1}\in(0.545,0.56)$.}
\end{figure}

The largest Lyapunov exponent, $\Lambda$, and its variance, $\mu$, that is the
variance of $\Lambda$ from finite time Lyapunov exponents $\lambda_i(N)$'s,
$i=1,2,\cdots,M$ of length $N$, defined as 
\begin{equation}
\mu=\frac{1}{M}\sum^{M}_{i=1}(\Lambda-\lambda_i(N))^2
\end{equation}
are shown in Figs.~\ref{fig12} in the range of $E_{1}\in(0.54,0.546)$. The
attractor depicted in Fig.~\ref{fig11}d(i) for the value of $E_{1}=0.546$ is
strange but it is nonchaotic as evidenced by the negative value of the Lyapunov
exponent shown in Fig.~\ref{fig12}a. It is also to be noted that both the
Lyapunov exponents and its variance (Fig.~\ref{fig12}b) clearly indicate a
critical value of amplitude $E_1^c=0.5432$, ($E_{1} < E_1^c$), below which 
torus exists and above which, ($E_{1} > E_1^c$),  SNA appears.  The regions
of torus and SNA are clearly indicated by smooth and irregular variations,
respectively, in the values of both the Lyapunov exponents $\Lambda$ and its
variance $\mu$. Finally the transition of SNA into a chaotic attractor is confirmed by the change in the largest Lyapunov exponent from negative to positive values at $E_1^c=0.55$ as shown in the inset of Fig.~\ref{fig12}a. 

\subsection{Spectral distribution function and scaling laws}
\begin{figure}
\centering
\includegraphics[width=1.0\columnwidth]{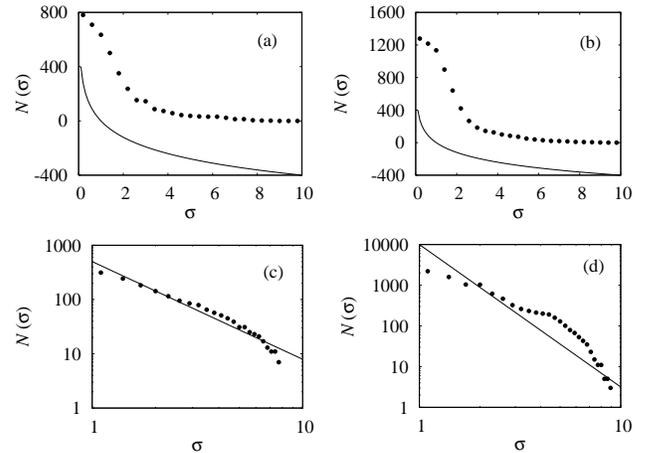}
\caption{\label{fig13} Spectral distribution function (filled circles)
calculated numerically. (a) torus (Fig.~\ref{fig11}a), (b) bubbled torus (Fig.~\ref{fig11}b), (c) SNA (Fig.~\ref{fig11}d) and (d) chaotic attractor (Fig.~\ref{fig11}e). Solid lines in (a) and (b) corresponds to the scaling relation
$N(\sigma)\sim\log_{10}(1/ \sigma)$ and in (c) and (d) correspond to the scaling
relation $N(\sigma)\sim\sigma^{-\beta}$, with$~ \beta=1.88$ and $3.5$, respectively.}
\end{figure}

In order to distinguish further whether the attractors depicted in Figs.~\ref{fig11} are
quasiperiodic or strange nonchaotic or chaotic attractors, we proceed to quantify the
changes in their power spectra.  The spectral distribution function, defined as
the number of peaks in the Fourier amplitude spectrum larger than some value
$\sigma$, is used to distinguish between  quasiperiodic attractors and SNA as well as SNAs and chaotic attractors. The quasiperiodic attractors obey a scaling relationship
$N(\sigma)\sim\log_{10}(1/ \sigma)$, while the SNAs satisfy a scaling power-law
relationship $N(\sigma)\sim\sigma^{-\beta}, 1<\beta<2$~\cite{fjreo1987}.  Similarly for the chaotic attractor, the scaling relation is $N(\sigma)\sim\sigma^{-\beta}, \beta>2$.
Spectral distribution functions (filled circles) of the torus 
(Fig.~\ref{fig11}a) and bubbled torus (Fig.~\ref{fig11}b) satisfy the scaling relation $N(\sigma)\sim\log_{10}(1/
\sigma)$ as indicated by the solid line in Figs.~\ref{fig13}a and \ref{fig13}b, respectively, which is the
characteristic of a torus. On the other hand the  spectral distribution function
of the SNA (Fig.~\ref{fig11}d) exhibits power-law behavior as depicted in
Fig.~\ref{fig13}c (filled circles) with the value of the exponent 
$\beta=1.88$,  confirming the existence of SNA. For the chaotic attractor (Fig.~\ref{fig11}e), the scaling exponent (Fig.~\ref{fig13}d) turns out to be $\beta=3.5$ as required.  Again the solid lines in 
Figs.~\ref{fig13}c and \ref{fig13}d  represent the scaling law for SNA and chaos, respectively. 

\subsection{Distribution of local Lyapunov exponents}
In addition to the qualitative discussion through the Poincar\'e surface of section plots in the
($\phi,x$)  plane  (Figs.~\ref{fig10} and \ref{fig10a}) in distinguishing the
type of route through which SNA appears, it is also possible to distinguish the
same  using the distribution of a quantitative measure, namely finite time
Lyapunov exponents.  It has been shown ~\cite{apvm1997} that a typical
trajectory on a SNA actually possesses positive Lyapunov exponents in finite
time intervals, although the asymptotic exponent is negative.  As  a
consequence, it is possible to observe different characteristics of SNAs created
through different mechanisms by a study of the differences in the distribution
of finite time  exponents $P(N,\lambda)$~\cite{apvm1997}.  The distribution can
be obtained by taking a long trajectory and dividing it into segments of length
$N$, from which the local Lyapunov exponent can be  calculated.  In the limit of
large $N$, this distribution will collapse to a $\delta$ function
$P(N,\lambda)\rightarrow\delta(\Lambda-\lambda)$.  The deviations from$-$and the
approach to$-$the limit can be very different for SNAs created through different
mechanisms~\cite{apvm1997}.

\begin{figure}
\centering
\includegraphics[width=1.0\columnwidth]{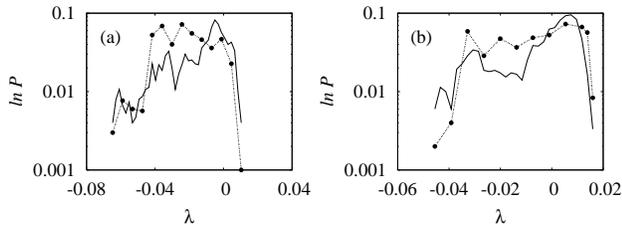}
\caption{\label{fig14} Distribution of finite time Lyapunov exponent calculated
from both numerical data (solid line) and experimental data (dashed line) of
(a) torus (Fig.~\ref{fig11}a) and (b) SNA (Fig.~\ref{fig11}d).}
\end{figure}

We have calculated the distribution of local Lyapunov exponents $P(N,\lambda)$, 
for $N=2000$, for the attractors shown in  Figs.~\ref{fig11}a(i)
and~\ref{fig11}d(i) in order to confirm the nature of transition to SNA. The distribution
of the local Lyapunov exponents for the period-3 torus (solid line) is shown in
Fig.~\ref{fig14}a in which the local Lyapunov exponents are peaked about the
largest Lyapunov exponents (negative values) of the torus while that of the SNA shown
in Fig.~\ref{fig14}b by solid line has its maximum at a positive value of the
local Lyapunov exponents. The distribution of local Lyapunov exponents for SNA 
exhibits an elongated tail in its  negative values because of the fact
that  in the bubbling transition parts of the strands of period-3 torus remain
unaffected even after the birth of SNA which
contributes largely to the negative values. This
confirms the existence of bubbling transition to strange nonchaotic attractor.

\section{Bubbling route to SNA: Experimental confirmation}

As a next step, in order to confirm the results of our numerical simulation in the
experimental circuit shown in Fig.~\ref{fig1}, 
a snapshot of the dynamical behavior for the corresponding values of the
experimental parameters is obtained (as mentioned in Sec.~\ref{cr}) and compared with that of the numerical
results. Further, the corresponding experimental data are analyzed using various
quantification measures mentioned in the previous section to confirm the nature of the dynamical
behavior.

\begin{figure}
\centering
\includegraphics[width=0.9\columnwidth]{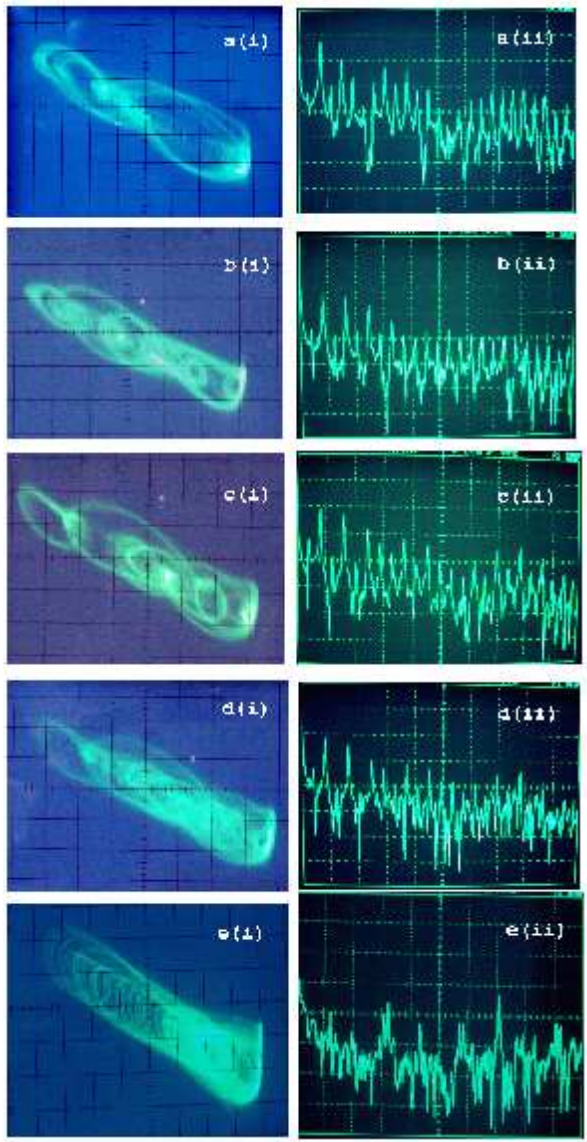}
\caption{\label{fig15} Snapshots of the experimental attractors and their power
spectrum of  the circuit shown in Fig.~\ref{fig1} for the corresponding values
of  the frequency $\omega_{f1}$ and the amplitude $E_{f1}$ of the sinusoidal
forcing in Fig.~\ref{fig10}. (a) period-3 torus (3T), (b) bubbled period-3
torus, (c) period-3 torus with enlarged bubbles, (d) fractalized bubbles
(SNA) and (e) chaotic attractor: (i) phase portrait in the $(v_C,i_L)$ space; (ii) power spectrum.}
\end{figure}

\subsection{Phase portraits and power spectra}
We have depicted the snapshots of the phase portraits and the corresponding
power spectra of the attractors as seen in the oscilloscope (which is connected
to the circuit shown in Fig.~\ref{fig1}) in Fig.~\ref{fig15} for the
corresponding values of the parameters of numerical simulation. Experimental
period-3 torus and its power spectrum corresponding to the numerical results,
Figs.~\ref{fig11}a, are shown in Figs.~\ref{fig15}a(i) and~\ref{fig15}a(ii). The
attractors in the bubbling regime for the values of the amplitude of the
sinusoidal forcing $E_{f1}=0.26V$ and $0.27V$ are shown in 
Figs.~\ref{fig15}b(i) and c(i), respectively. The corresponding power spectra
are shown in Figs.~\ref{fig15}b(ii) and c(ii), respectively. Experimental phase portrait of
the strange nonchaotic attractor and its power spectrum for the value of
$E_{f1}=0.273V$ are depicted in Figs.~\ref{fig15}d(i) and~\ref{fig15}d(ii),
respectively.  It is also seen that the spectra of the quasiperiodic attractors
are concentrated at a small discrete set  of frequencies while the spectrum  of
the  SNA has a much richer set of  harmonics. Further the resemblance of the
attractors illustrated in Figs.~\ref{fig15}(i) with that of the attractors in 
Figs.~\ref{fig11}(i) confirms the existence of bubbling transition to SNA in
this negative conductance series LCR circuit with diode having both the
sinusoidal and nonsinusoidal forces as quasiperiodic forcings.  Finally, the chaotic attractor for $E_{f1}=0.28V$ and its power spectrum are shown in Figs.~\ref{fig15}e. 

\subsection{Spectral distribution function and scaling laws}
\begin{figure}
\centering
\includegraphics[width=1.0\columnwidth]{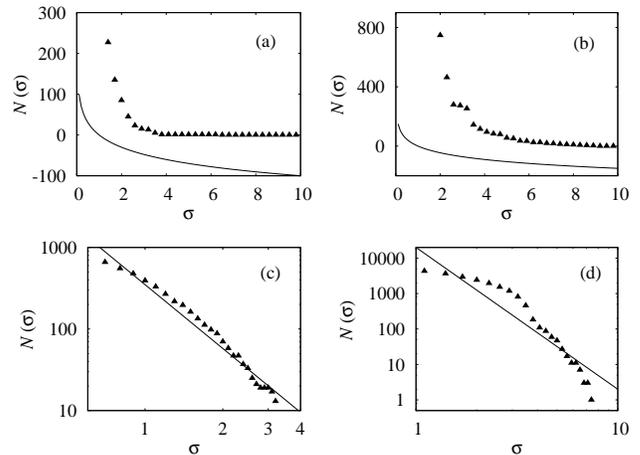}
\caption{\label{fig16} Spectral distribution function (filled triangles)
calculated from the experimental data of (a) torus (Fig.~\ref{fig15}a), (b) bubbled torus (Fig.~\ref{fig15}b), (c) SNA (Fig.~\ref{fig15}d) and (d) chaotic attractor (Fig.~\ref{fig15}e). Solid curve/line in (a) and (b) correspond to the scaling
relationship for the quasiperiodic attractors and in (c) and (d) correspond to the scaling relation for the SNA and chaotic attractor, respectively.}
\end{figure}

In order to confirm that the experimental phase portraits shown in 
Figs.~\ref{fig15}a, \ref{fig15}b, \ref{fig15}d and \ref{fig15}e are indeed that of torus, bubbled torus, SNA and chaotic attractor, respectively, the corresponding data are examined for the behavior
in their spectral distribution.  Figs.~\ref{fig16}a and \ref{fig16}b show the spectral
distribution function (filled triangles) for the torus in Figs.~\ref{fig15}a and \ref{fig15}b
satisfying the scaling relation $N(\sigma)\sim\log_{10}(1/ \sigma)$ as indicated
by the solid lines while that of the SNA (Fig.~\ref{fig15}d) shown in
Fig.~\ref{fig16}c obey power-law distribution with the value of the exponent
$\beta=1.96$ lying within the characteristic value for SNAs.  For the chaotic attractor (Fig.~\ref{fig15}e), the scaling exponent turns out to be $\beta=4.0$ (Fig.~\ref{fig16}d) as expected. 

\subsection{Local Lyapunov exponents}
Further, in order to examine
whether the SNA shown in Fig.~\ref{fig15}d arises from the bubbling transition,
the distribution of the local Lyapunov exponents calculated from the
experimental data of the torus (Fig.~\ref{fig15}a) and the SNA
(Fig.~\ref{fig15}d) are depicted in Figs.~\ref{fig14}a and \ref{fig14}b
respectively as dashed lines. The elongated tail  in the distribution of the
local Lyapunov exponents even for SNA (Fig.~\ref{fig15}d) confirms the existence
of undisturbed strands as shown in Fig.~\ref{fig10}d, thereby confirming
experimentally the birth of SNA via the bubbling transition.

\section{Summary and conclusion} 

In this paper,  we have reported the birth of strange nonchaotic attractors
through a novel route which we term as the \emph{bubbling route to SNA} in a
negative conductance series LCR circuit with the diode containing nonsinusoidal
(square wave) force as one of the quasiperiodic forcings.  At first, we have
presented the numerical analysis of the dynamical system, namely~Eq.~(\ref{nor_eq}) of
the circuit (Fig.~\ref{fig1}) for suitable ranges of the amplitude $E_{1}$ and
the frequency $\omega_{1}$ of the sinusoidal force while the other parameters
are held fixed.  Following this, we have also confirmed the numerical results
experimentally by the snapshots of the  phase portraits of the quasiperiodic
attractors and SNAs as well as chaotic attractors for the corresponding values of the circuit parameters.
Further, the numerical and experimental data have been analyzed using various
quantification measures attributing to the existence of torus, SNA, birth of
SNA through the bubbling route and transition to chaos.  In particular, we have characterized the
quasiperiodic attractors, SNAs and chaotic attractors using maximal Lyapunov exponent and its
variance, Poincar\'e maps, Fourier amplitude spectra, spectral distribution
function and distribution of finite time Lyapunov exponents.  The distribution
of local Lyapunov exponents indeed clearly distinguishes the characteristic
properties of both the torus and the SNA,  confirming the existence of bubbling
route to the SNA. The experimental observations,  numerical simulations and
characteristic analysis showed that the simple dissipative  negative conductance
series LCR circuit even with a nonsinusoidal (square wave) force as one of the 
quasiperiodic forces does indeed admit strange nonchaotic behaviors of 
different  types and in particular admits a novel bubbling route to SNA.

\begin{acknowledgments}
This work has been supported by a Department of Science and Technology,
Government of India sponsored IRHPA research project. The work of M. L. has also
been supported by a DST Ramanna Fellowhip research grant.
\end{acknowledgments}

\appendix 
\section{\label{a1} Identification and characterization of SNAs and their 
routes} 

Torus, SNA and chaotic attractors and the transitions between them through
different routes can be identified and characterized through various qualitative
and quantitative measures.  In this Appendix, we summarize the main measures
used in  the recent
literature~\cite{avml2000,avml1997,apvm1997,avml2001,csztlc1997,sykwl2004,tkloc1997,
tykb1997,avkm1999,wldmls1990,ktdvsk2006} in the analysis of transitions to SNAs from torus
attractors and from SNAs to chaotic attractors. In the present work also, we utilize these measures.

\begin{enumerate}
\item \emph{Qualitative measures:}\\

Geometrically smooth (torus) and non-smooth (SNAs and chaotic) attractors can
be  distinguished qualitatively using Poincar\'e surface of sections and Fourier
spectra. The Poincar\'e surface of section shows smooth strands for quasiperiodic
attractors, non-smooth strands for SNAs, widely interspersed points throughout
the phase space for chaotic attractors, which clearly reveals whether an
attractor possesses a geometrically smooth or complicated structure.  The
spectra of the quasiperiodic attractors are concentrated at a small discrete set
of frequencies while the spectra of SNAs and chaotic attractors have a much
richer set of harmonics.

Further, different types of routes to SNAs and their mechanisms for their
formation  can also be identified qualitatively using the Poincar\'e surface of
sections by observing the nature of the dynamics  in these plots as a function
of the control parameter. Different routes for the formation of SNAs have
different characteristic dynamics in their Poincar\'e surface of section.

\item \emph{Quantative measures:}
\begin{enumerate}

\item The largest Lyapunov exponents can be used to distinguish between (i)
torus and SNAs and (ii) SNAs  and chaotic attractors. Torus motion is
characterized by a smooth negative Lyapunov exponent, SNAs are characterized by
either zero or non-smooth negative Lyapunov exponents as a function of control
parameters and chaotic attractors have atleast one  positive Lyapunov exponent.
Further, the transition from torus to  SNAs exhibits different  signatures in
the values of the largest Lyapunov exponents and their variance for different
routes  to SNAs~\cite{avml2001}.

\item Further, torus and SNA can also be distinguished quantitatively by using
the spectral distribution function, which is defined as the number of peaks in
the Fourier amplitude spectrum larger than some value $\sigma$~\cite{abeo1985}. The
quasiperiodic attractors obey a scaling relationship $N(\sigma)\sim\log_{10}(1/
\sigma)$, while the SNAs satisfy a scaling power-law relationship
$N(\sigma)\sim\sigma^{-\beta}, 1<\beta<2$.  For chaos, the scaling exponent $\beta>2$.

\item Finer distinction between the different types of routes for the formation
of SNAs can also made using the distribution of finite time Lyapunov exponents.
Different routes are  characterized by different types of the distribution of
finite time Lyapunov exponents~\cite{apvm1997}.

\end{enumerate}

\end{enumerate}

The different signatures of the above quantitative measures corresponding to
different scenarios (routes) for the formation of three well known types of SNAs
are tabulated in Table~II.

\begin{widetext}

\begin{table}
\caption{Different signatures of the largest Lyapunov exponents and its
variance, and the distribution of finite time Lyapunov exponents for  the
formation of three prominent types of SNAs}
%\small
%\begin{tabular}{|p{7cm}|p{7cm}|}
%\hline
%{\bf \emph \;\;\;\;\;\;\;\;\;\;\;\;\;\;Type of route} & {\bf \emph \;\;\;\;\;\;\;\;\;\;\;\;\;\;\;\;Mechanism}  \\
\begin{ruledtabular}
\begin{tabular}{p{3cm}p{5cm}p{4cm}p{8cm}}
{\bf \emph \;\;Type of route} & {\bf \emph Lyapunov exponent $\Lambda$} & {\bf
\emph \;\;\;\; Variance $\mu$} & {\bf \emph \;Distribution of finite time } \\
 & &  & {\bf \emph \;Lyapunov  exponents $P(N,\lambda)$ } \\
\hline

Heagy-Hammel~\cite{jfhsmh1994} & Irregular in the SNA region  and& Small in torus & Distribution
shifts continuously to \\ 
& smooth  in the torus region&and large in
SNA&larger  exponents but the shape differs \\ 
&&& for torus and SNA\\

Gradual & Increases  slowly during the & Increases only slowly  & 
Distribution shifts continuously to\\
fractalization~\cite{tnkk1996}& transition from torus to SNA && larger exponents but the shape\\
&&& remains the same for torus and SNA\\

Intermittency\cite{apvm1997,avkm1999}& Abrupt change during the & Abrupt change  at the& 
Stretched exponential  tail  and  \\
& transition from torus to SNA & transition  point&asymmetric distribution\\

%\hline
%\hline
%\end{tabular}
\end{tabular}
\end{ruledtabular}
\end{table} 
\end{widetext}

\newpage %Just because of unusual number of tables stacked at end


\begin{thebibliography}{49}
\expandafter\ifx\csname natexlab\endcsname\relax\def\natexlab#1{#1}\fi
\expandafter\ifx\csname bibnamefont\endcsname\relax
  \def\bibnamefont#1{#1}\fi
\expandafter\ifx\csname bibfnamefont\endcsname\relax
  \def\bibfnamefont#1{#1}\fi
\expandafter\ifx\csname citenamefont\endcsname\relax
  \def\citenamefont#1{#1}\fi
\expandafter\ifx\csname url\endcsname\relax
  \def\url#1{\texttt{#1}}\fi
\expandafter\ifx\csname urlprefix\endcsname\relax\def\urlprefix{URL }\fi
\providecommand{\bibinfo}[2]{#2}
\providecommand{\eprint}[2][]{\url{#2}}

\bibitem[{\citenamefont{Grebogi et~al.}(1984)\citenamefont{Grebogi, Ott,
  Pelikan, and Yorke}}]{cgeo1984}
\bibinfo{author}{\bibfnamefont{C.}~\bibnamefont{Grebogi}},
  \bibinfo{author}{\bibfnamefont{E.}~\bibnamefont{Ott}},
  \bibinfo{author}{\bibfnamefont{S.}~\bibnamefont{Pelikan}}, \bibnamefont{and}
  \bibinfo{author}{\bibfnamefont{J.~A.} \bibnamefont{Yorke}},
  \bibinfo{journal}{Physica D} \textbf{\bibinfo{volume}{13}},
  \bibinfo{pages}{261} (\bibinfo{year}{1984}).

\bibitem[{\citenamefont{Romeiras and Ott}(1987)}]{fjreo1987}
\bibinfo{author}{\bibfnamefont{F.~J.} \bibnamefont{Romeiras}} \bibnamefont{and}
  \bibinfo{author}{\bibfnamefont{E.}~\bibnamefont{Ott}},
  \bibinfo{journal}{Phys. Rev. A} \textbf{\bibinfo{volume}{35}},
  \bibinfo{pages}{4404} (\bibinfo{year}{1987});
%  [{\citenamefont{Romeiras et~al.}(1987)\citenamefont{Romeiras, Bondeson,
%  Ott, Jr., and Grebogi}}]
\bibinfo{author}{\bibfnamefont{F.~J.} \bibnamefont{Romeiras}},
  \bibinfo{author}{\bibfnamefont{A.}~\bibnamefont{Bondeson}},
  \bibinfo{author}{\bibfnamefont{E.}~\bibnamefont{Ott}},
  \bibinfo{author}{\bibfnamefont{T.~M.}~\bibnamefont{Andonsen, Jr.}},
  \bibnamefont{and} \bibinfo{author}{\bibfnamefont{C.}~\bibnamefont{Grebogi}},
  \bibinfo{journal}{Physica D} \textbf{\bibinfo{volume}{26}},
  \bibinfo{pages}{277} (\bibinfo{year}{1987}).


\bibitem[{\citenamefont{Bondeson et~al.}(1985)\citenamefont{Bondeson, Ott, and
  Jr.}}]{abeo1985}
\bibinfo{author}{\bibfnamefont{A.}~\bibnamefont{Bondeson}},
  \bibinfo{author}{\bibfnamefont{E.}~\bibnamefont{Ott}}, \bibnamefont{and}
  \bibinfo{author}{\bibfnamefont{T.~M.}~\bibnamefont{Antonsen, Jr.}},
  \bibinfo{journal}{Phys. Rev. Lett.} \textbf{\bibinfo{volume}{55}},
  \bibinfo{pages}{2103} (\bibinfo{year}{1985});
\bibinfo{author}{\bibfnamefont{Y.~C.} \bibnamefont{Lai}},
  \bibinfo{journal}{Phys. Rev. E} \textbf{\bibinfo{volume}{53}},
  \bibinfo{pages}{57} (\bibinfo{year}{1996}).

\bibitem[{\citenamefont{Ding et~al.}(1989)\citenamefont{Ding, Grebogi, and
  Ott}}]{mdcg1989}
\bibinfo{author}{\bibfnamefont{M.}~\bibnamefont{Ding}},
  \bibinfo{author}{\bibfnamefont{C.}~\bibnamefont{Grebogi}}, \bibnamefont{and}
  \bibinfo{author}{\bibfnamefont{E.}~\bibnamefont{Ott}},
  \bibinfo{journal}{Phys. Rev. A} \textbf{\bibinfo{volume}{39}},
  \bibinfo{pages}{2593} (\bibinfo{year}{1989});
\bibinfo{author}{\bibfnamefont{M.}~\bibnamefont{Ding}} \bibnamefont{and}
  \bibinfo{author}{\bibfnamefont{J.~A.}~\bibnamefont{Scott Relso}},
  \bibinfo{journal}{Int. J. Bifurcation and Chaos Appl. Sci. Eng.}
  \textbf{\bibinfo{volume}{4}}, \bibinfo{pages}{533} (\bibinfo{year}{1994}).

\bibitem[{\citenamefont{Heagy and Ditto}(1991)}]{jfhwld1991}
\bibinfo{author}{\bibfnamefont{J.~F.} \bibnamefont{Heagy}} \bibnamefont{and}
  \bibinfo{author}{\bibfnamefont{W.~L.} \bibnamefont{Ditto}},
  \bibinfo{journal}{J. Nonlinear Sci.} \textbf{\bibinfo{volume}{1}},
  \bibinfo{pages}{423} (\bibinfo{year}{1991});
\bibinfo{author}{\bibfnamefont{J.~I.} \bibnamefont{Staglino}},
  \bibinfo{author}{\bibfnamefont{J.~M.} \bibnamefont{Wersinger}},
  \bibnamefont{and} \bibinfo{author}{\bibfnamefont{E.~E.}
  \bibnamefont{Slaminka}}, \bibinfo{journal}{Physica D}
  \textbf{\bibinfo{volume}{92}}, \bibinfo{pages}{164} (\bibinfo{year}{1996}).

\bibitem[{\citenamefont{Yalcinkaya and Lai}(1996)}]{tyycl1996}
\bibinfo{author}{\bibfnamefont{T.}~\bibnamefont{Yalcinkaya}} \bibnamefont{and}
  \bibinfo{author}{\bibfnamefont{Y.~C.} \bibnamefont{Lai}},
  \bibinfo{journal}{Phys. Rev. Lett.} \textbf{\bibinfo{volume}{77}},
  \bibinfo{pages}{5039} (\bibinfo{year}{1996}).

\bibitem[{\citenamefont{Kapitaniak and Wojewoda}(1993)}]{tkjw1993}
\bibinfo{author}{\bibfnamefont{T.}~\bibnamefont{Kapitaniak}} \bibnamefont{and}
  \bibinfo{author}{\bibfnamefont{J.}~\bibnamefont{Wojewoda}},
  \emph{\bibinfo{title}{Attractors of Quasiperiodically Forced Systems}}
  (\bibinfo{publisher}{World Scientific}, \bibinfo{address}{singapore},
  \bibinfo{year}{1993}).

\bibitem[{\citenamefont{Venkatesan et~al.}(2000)\citenamefont{Venkatesan,
  Lakshmanan, Prasad, and Ramaswamy}}]{avml2000}
\bibinfo{author}{\bibfnamefont{A.}~\bibnamefont{Venkatesan}},
  \bibinfo{author}{\bibfnamefont{M.}~\bibnamefont{Lakshmanan}},
  \bibinfo{author}{\bibfnamefont{A.}~\bibnamefont{Prasad}}, \bibnamefont{and}
  \bibinfo{author}{\bibfnamefont{R.}~\bibnamefont{Ramaswamy}},
  \bibinfo{journal}{Phys. Rev. E} \textbf{\bibinfo{volume}{61}},
  \bibinfo{pages}{3641} (\bibinfo{year}{2000}).

\bibitem[{\citenamefont{Venkatesan and Lakshmanan}(1997)}]{avml1997}
\bibinfo{author}{\bibfnamefont{A.}~\bibnamefont{Venkatesan}} \bibnamefont{and}
  \bibinfo{author}{\bibfnamefont{M.}~\bibnamefont{Lakshmanan}},
  \bibinfo{journal}{Phys. Rev. E} \textbf{\bibinfo{volume}{55}},
  \bibinfo{pages}{5134} (\bibinfo{year}{1997});
\bibinfo{author}{\bibfnamefont{A.}~\bibnamefont{Venkatesan}} \bibnamefont{and}
  \bibinfo{author}{\bibfnamefont{M.}~\bibnamefont{Lakshmanan}},
  \bibinfo{journal}{Phys. Rev. E} \textbf{\bibinfo{volume}{58}},
  \bibinfo{pages}{3008} (\bibinfo{year}{1998}).

\bibitem[{\citenamefont{Prasad et~al.}(1997)\citenamefont{Prasad, Mehra, and
  Ramaswamy}}]{apvm1997}
\bibinfo{author}{\bibfnamefont{A.}~\bibnamefont{Prasad}},
  \bibinfo{author}{\bibfnamefont{V.}~\bibnamefont{Mehra}}, \bibnamefont{and}
  \bibinfo{author}{\bibfnamefont{R.}~\bibnamefont{Ramaswamy}},
  \bibinfo{journal}{Phys. Rev. Lett.} \textbf{\bibinfo{volume}{79}},
  \bibinfo{pages}{4127} (\bibinfo{year}{1997});
  \bibinfo{journal}{Phys. Rev. E} \textbf{\bibinfo{volume}{57}},
  \bibinfo{pages}{1576} (\bibinfo{year}{1998}).

\bibitem[{\citenamefont{Pikovsky and Feudel}(1995)}]{aspuf1995}
\bibinfo{author}{\bibfnamefont{A.~S.} \bibnamefont{Pikovsky}} \bibnamefont{and}
  \bibinfo{author}{\bibfnamefont{U.}~\bibnamefont{Feudel}},
  \bibinfo{journal}{Chaos} \textbf{\bibinfo{volume}{5}}, \bibinfo{pages}{253}
  (\bibinfo{year}{1995});
\bibinfo{author}{\bibfnamefont{U.}~\bibnamefont{Feudel}},
  \bibinfo{author}{\bibfnamefont{J.}~\bibnamefont{Kurths}}, \bibnamefont{and}
  \bibinfo{author}{\bibfnamefont{A.~S.} \bibnamefont{Pikovsky}},
  \bibinfo{journal}{Physica D} \textbf{\bibinfo{volume}{88}},
  \bibinfo{pages}{176} (\bibinfo{year}{1995});
%\bibitem[{\citenamefont{Pikovsky and Feudel}(1994)}]{aspuf1994}
\bibinfo{author}{\bibfnamefont{A.~S.} \bibnamefont{Pikovsky}} \bibnamefont{and}
  \bibinfo{author}{\bibfnamefont{U.}~\bibnamefont{Feudel}},
  \bibinfo{journal}{J. Phys. A} \textbf{\bibinfo{volume}{27}},
  \bibinfo{pages}{5209} (\bibinfo{year}{1994});
\bibinfo{author}{\bibfnamefont{S.~P.} \bibnamefont{Kuznetsov}},
  \bibinfo{author}{\bibfnamefont{A.~S.} \bibnamefont{Pikovsky}},
  \bibnamefont{and} \bibinfo{author}{\bibfnamefont{U.}~\bibnamefont{Feudel}},
  \bibinfo{journal}{Phys. Rev. E} \textbf{\bibinfo{volume}{51}},
  \bibinfo{pages}{R1629} (\bibinfo{year}{1995});
\bibinfo{author}{\bibfnamefont{A.}~\bibnamefont{Witt}},
  \bibinfo{author}{\bibfnamefont{U.}~\bibnamefont{Feudel}}, \bibnamefont{and}
  \bibinfo{author}{\bibfnamefont{A.~S.} \bibnamefont{Pikovsky}},
  \bibinfo{journal}{Physica D} \textbf{\bibinfo{volume}{109}},
  \bibinfo{pages}{180} (\bibinfo{year}{1997}).

\bibitem[{\citenamefont{Anishchensko et~al.}(1996)\citenamefont{Anishchenko,
  Vadivasova, and Sosnovtseva}}]{vsatev1996}
\bibinfo{author}{\bibfnamefont{V.~S.} \bibnamefont{Anishchenko}},
  \bibinfo{author}{\bibfnamefont{T.~E.} \bibnamefont{Vadivasova}},
  \bibnamefont{and}
  \bibinfo{author}{\bibfnamefont{O.}~\bibnamefont{Sosnovtseva}},
  \bibinfo{journal}{Phys. Rev. E} \textbf{\bibinfo{volume}{53}},
  \bibinfo{pages}{4451} (\bibinfo{year}{1996});
\bibinfo{author}{\bibfnamefont{O.}~\bibnamefont{Sosnovtseva}},
  \bibinfo{author}{\bibfnamefont{U.}~\bibnamefont{Feudel}},
  \bibinfo{author}{\bibfnamefont{J.}~\bibnamefont{Kurths}}, \bibnamefont{and}
  \bibinfo{author}{\bibfnamefont{A.~S.} \bibnamefont{Pikovsky}},
  \bibinfo{journal}{Phys. Lett. A} \textbf{\bibinfo{volume}{218}},
  \bibinfo{pages}{225} (\bibinfo{year}{1996});
\bibinfo{author}{\bibfnamefont{S.} \bibnamefont{Kuznetsov}},
  \bibinfo{author}{\bibfnamefont{U.}~\bibnamefont{Feudel}}, \bibnamefont{and}
  \bibinfo{author}{\bibfnamefont{A.} \bibnamefont{Pikovsky}},
  \bibinfo{journal}{Phys. Rev. E} \textbf{\bibinfo{volume}{57}},
  \bibinfo{pages}{1585} (\bibinfo{year}{1998}).

\bibitem[{\citenamefont{Nishikawa and Kaneko}(1996)}]{tnkk1996}
\bibinfo{author}{\bibfnamefont{K.}~\bibnamefont{Kaneko}},
  \bibinfo{journal}{Pro. Theor. Phys.} \textbf{\bibinfo{volume}{71}},
  \bibinfo{pages}{140} (\bibinfo{year}{1984});
\bibinfo{author}{\bibfnamefont{T.}~\bibnamefont{Nishikawa}} \bibnamefont{and}
  \bibinfo{author}{\bibfnamefont{K.}~\bibnamefont{Kaneko}},
  \bibinfo{journal}{Phys. Rev. E} \textbf{\bibinfo{volume}{54}},
  \bibinfo{pages}{6114} (\bibinfo{year}{1996}).

\bibitem[{\citenamefont{Venkatesan and Lakshmanan}(2001)}]{avml2001}
\bibinfo{author}{\bibfnamefont{A.}~\bibnamefont{Venkatesan}} \bibnamefont{and}
  \bibinfo{author}{\bibfnamefont{M.}~\bibnamefont{Lakshmanan}},
  \bibinfo{journal}{Phys. Rev. E} \textbf{\bibinfo{volume}{63}},
  \bibinfo{pages}{026219} (\bibinfo{year}{2001}).

\bibitem[{\citenamefont{Hunt and Ott}(2001)}]{brheo2001}
\bibinfo{author}{\bibfnamefont{B.~R.} \bibnamefont{Hunt}} \bibnamefont{and}
  \bibinfo{author}{\bibfnamefont{E.}~\bibnamefont{Ott}},
  \bibinfo{journal}{Phys. Rev. Lett.} \textbf{\bibinfo{volume}{87}},
  \bibinfo{pages}{254101} (\bibinfo{year}{2001});
\bibinfo{author}{\bibfnamefont{J.~W.} \bibnamefont{Kim}},
  \bibinfo{author}{\bibfnamefont{S.~Y.} \bibnamefont{Kim}},
  \bibinfo{author}{\bibfnamefont{B.~R.} \bibnamefont{Hunt}}, \bibnamefont{and}
  \bibinfo{author}{\bibfnamefont{E.}~\bibnamefont{Ott}},
  \bibinfo{journal}{Phys. Rev. E} \textbf{\bibinfo{volume}{67}},
  \bibinfo{pages}{036211} (\bibinfo{year}{2003}{\natexlab{a}});
\bibinfo{author}{\bibfnamefont{S.~Y.} \bibnamefont{Kim}},
  \bibinfo{author}{\bibfnamefont{W.}~\bibnamefont{Lim}}, \bibnamefont{and}
  \bibinfo{author}{\bibfnamefont{E.}~\bibnamefont{Ott}},
  \bibinfo{journal}{Phys. Rev. E} \textbf{\bibinfo{volume}{67}},
  \bibinfo{pages}{056203} (\bibinfo{year}{2003}{\natexlab{b}});
\bibinfo{author}{\bibfnamefont{W.}~\bibnamefont{Lim}} \bibnamefont{and}
  \bibinfo{author}{\bibfnamefont{S.~Y.} \bibnamefont{Kim}},
  \bibinfo{journal}{J. Korean Physical Society} \textbf{\bibinfo{volume}{3}},
  \bibinfo{pages}{514} (\bibinfo{year}{2004}).

\bibitem[{\citenamefont{Heagy and Hammel}(1994)}]{jfhsmh1994}
\bibinfo{author}{\bibfnamefont{J.~F.} \bibnamefont{Heagy}} \bibnamefont{and}
  \bibinfo{author}{\bibfnamefont{S.~M.} \bibnamefont{Hammel}},
  \bibinfo{journal}{Physica D} \textbf{\bibinfo{volume}{70}},
  \bibinfo{pages}{140} (\bibinfo{year}{1994}).

\bibitem[{\citenamefont{Prasad et~al.}(1999)\citenamefont{Prasad, Ramaswamy,
  Satija, and Shah}}]{aprr1999}
\bibinfo{author}{\bibfnamefont{A.}~\bibnamefont{Prasad}},
  \bibinfo{author}{\bibfnamefont{R.}~\bibnamefont{Ramaswamy}},
  \bibinfo{author}{\bibfnamefont{I.~I}~\bibnamefont{Satija}}, \bibnamefont{and}
  \bibinfo{author}{\bibfnamefont{N.}~\bibnamefont{Shah}},
  \bibinfo{journal}{Phys. Rev. Lett} \textbf{\bibinfo{volume}{83}},
  \bibinfo{pages}{4530} (\bibinfo{year}{1999}).

\bibitem[{\citenamefont{Zhou and Chen}(1997)}]{csztlc1997}
\bibinfo{author}{\bibfnamefont{C.~S.} \bibnamefont{Zhou}} \bibnamefont{and}
  \bibinfo{author}{\bibfnamefont{T.~L.} \bibnamefont{Chen}},
  \bibinfo{journal}{Europhys. Lett.} \textbf{\bibinfo{volume}{38}},
  \bibinfo{pages}{261} (\bibinfo{year}{1997});
\bibinfo{author}{\bibfnamefont{R.}~\bibnamefont{Ramaswamy}},
  \bibinfo{journal}{Phys. Rev. E.} \textbf{\bibinfo{volume}{56}},
  \bibinfo{pages}{7294} (\bibinfo{year}{1997});
\bibinfo{author}{\bibfnamefont{R.}~\bibnamefont{Chacon}} \bibnamefont{and}
  \bibinfo{author}{\bibfnamefont{A.~M.} \bibnamefont{Gracia-Hoz}},
  \bibinfo{journal}{Europhys. Lett.} \textbf{\bibinfo{volume}{57}},
  \bibinfo{pages}{7} (\bibinfo{year}{2002}).

\bibitem[{\citenamefont{Kim and Lim}(2004)}]{sykwl2004}
\bibinfo{author}{\bibfnamefont{S.~Y.} \bibnamefont{Kim}} \bibnamefont{and}
  \bibinfo{author}{\bibfnamefont{W.}~\bibnamefont{Lim}}, \bibinfo{journal}{J.
  Phys. A} \textbf{\bibinfo{volume}{37}}, \bibinfo{pages}{6477}
  (\bibinfo{year}{2004}).

\bibitem[{\citenamefont{Kapitaniak et~al.}(1997)\citenamefont{Kapitaniak,
  and Chua}}]{tkloc1997}
\bibinfo{author}{\bibfnamefont{T.}~\bibnamefont{Kapitaniak}} \bibnamefont{and}
  \bibinfo{author}{\bibfnamefont{L.~O.} \bibnamefont{Chua}},
  \bibinfo{journal}{Int. J. Bifurcation and Chaos Appl. Sci. Eng.}
  \textbf{\bibinfo{volume}{7}}, \bibinfo{pages}{423} (\bibinfo{year}{1997}).

\bibitem[{\citenamefont{Yang and Bilimgut}(1997)}]{tykb1997}
\bibinfo{author}{\bibfnamefont{T.}~\bibnamefont{Yang}} \bibnamefont{and}
  \bibinfo{author}{\bibfnamefont{K.}~\bibnamefont{Bilimgut}},
  \bibinfo{journal}{Phys. Lett. A} \textbf{\bibinfo{volume}{236}},
  \bibinfo{pages}{494} (\bibinfo{year}{1997});
%\bibitem[{\citenamefont{Liu and Zhua}(1996)}]{zlzz1996}
\bibinfo{author}{\bibfnamefont{Z.}~\bibnamefont{Liu}} \bibnamefont{and}
  \bibinfo{author}{\bibfnamefont{Z.}~\bibnamefont{Zhua}},
  \bibinfo{journal}{Int. J. Bifurcation and Chaos Appl. Sci. Eng.}
  \textbf{\bibinfo{volume}{6}}, \bibinfo{pages}{1383} (\bibinfo{year}{1996});
\bibinfo{author}{\bibfnamefont{Z.}~\bibnamefont{Zhua}} \bibnamefont{and}
  \bibinfo{author}{\bibfnamefont{Z.}~\bibnamefont{Liu}}, \bibinfo{journal}{ibid}
  \textbf{\bibinfo{volume}{7}}, \bibinfo{pages}{227}
  (\bibinfo{year}{1997}).

\bibitem[{\citenamefont{Venkatesan et~al.}(1999)\citenamefont{Venkatesan,
  Murali, and Lakshmanan}}]{avkm1999}
\bibinfo{author}{\bibfnamefont{A.}~\bibnamefont{Venkatesan}},
  \bibinfo{author}{\bibfnamefont{K.}~\bibnamefont{Murali}}, \bibnamefont{and}
  \bibinfo{author}{\bibfnamefont{M.}~\bibnamefont{Lakshmanan}},
  \bibinfo{journal}{Phys. Lett. A} \textbf{\bibinfo{volume}{259}},
  \bibinfo{pages}{246} (\bibinfo{year}{1999}).

\bibitem[{\citenamefont{Ditto et~al.}(1990)\citenamefont{Ditto, Spano, Savage,
  Rauseo, Heagy, and Ott}}]{wldmls1990}
\bibinfo{author}{\bibfnamefont{W.~L.} \bibnamefont{Ditto}},
  \bibinfo{author}{\bibfnamefont{M.~L.} \bibnamefont{Spano}},
  \bibinfo{author}{\bibfnamefont{H.~T.} \bibnamefont{Savage}},
  \bibinfo{author}{\bibfnamefont{S.~N.} \bibnamefont{Rauseo}},
  \bibinfo{author}{\bibfnamefont{J.~F.} \bibnamefont{Heagy}}, \bibnamefont{and}
  \bibinfo{author}{\bibfnamefont{E.}~\bibnamefont{Ott}},
  \bibinfo{journal}{Phys. Rev. Lett.} \textbf{\bibinfo{volume}{65}},
  \bibinfo{pages}{533} (\bibinfo{year}{1990});
\bibinfo{author}{\bibfnamefont{T.}~\bibnamefont{Zhou}},
  \bibinfo{author}{\bibfnamefont{F.}~\bibnamefont{Moss}}, \bibnamefont{and}
  \bibinfo{author}{\bibfnamefont{A.}~\bibnamefont{Bulsara}},
  \bibinfo{journal}{Phys. Rev. A} \textbf{\bibinfo{volume}{45}},
  \bibinfo{pages}{5394} (\bibinfo{year}{1992});
\bibinfo{author}{\bibfnamefont{W.~X.} \bibnamefont{Ding}},
  \bibinfo{author}{\bibfnamefont{H.}~\bibnamefont{Deutsch}},
  \bibinfo{author}{\bibfnamefont{A.}~\bibnamefont{Dinklage}},
  \bibnamefont{and} \bibinfo{author}{\bibfnamefont{C.}~\bibnamefont{Wilke}},
  \bibinfo{journal}{Phys. Rev. E} \textbf{\bibinfo{volume}{55}},
  \bibinfo{pages}{3769} (\bibinfo{year}{1997});
\bibinfo{author}{\bibfnamefont{J.~A.} \bibnamefont{Ketoja}} \bibnamefont{and}
  \bibinfo{author}{\bibfnamefont{I.}~\bibnamefont{Satija}},
  \bibinfo{journal}{Physica D} \textbf{\bibinfo{volume}{109}},
  \bibinfo{pages}{70} (\bibinfo{year}{1997}).


\bibitem[{\citenamefont{Thamilmaran et~al.}(2006)\citenamefont{Thamilmaran,
  Senthilkumar, Venkatesan, and Lakshmanan}}]{ktdvsk2006}
\bibinfo{author}{\bibfnamefont{K.}~\bibnamefont{Thamilmaran}},
  \bibinfo{author}{\bibfnamefont{D.~V.} \bibnamefont{Senthilkumar}},
  \bibinfo{author}{\bibfnamefont{A.}~\bibnamefont{Venkatesan}},
  \bibnamefont{and} \bibinfo{author}{\bibfnamefont{M.}~\bibnamefont{Lakshmanan}},
  \bibinfo{journal}{Phys. Rev. E.}
  \textbf{\bibinfo{volume}{74}}, \bibinfo{pages}{036205} (\bibinfo{year}{2006}).


\bibitem[{\citenamefont{Kapitaniak et~al.}(1997)\citenamefont{Kapitaniak,
  and Chua}}]{apssn2001}
  \bibinfo{author}{\bibfnamefont{A.}~\bibnamefont{Prasad}},
\bibinfo{author}{\bibfnamefont{S. S.}~\bibnamefont{Negi}} \bibnamefont{and}
  \bibinfo{author}{\bibfnamefont{R.} \bibnamefont{Ramaswamy}},
  \bibinfo{journal}{Int. J. Bifurcation and Chaos Appl. Sci. Eng.}
  \textbf{\bibinfo{volume}{11}}, \bibinfo{pages}{291} (\bibinfo{year}{2001});
    \bibinfo{author}{\bibfnamefont{A.}~\bibnamefont{Prasad}},
\bibinfo{author}{\bibfnamefont{A.}~\bibnamefont{Nandi}} \bibnamefont{and}
  \bibinfo{author}{\bibfnamefont{R.} \bibnamefont{Ramaswamy}},
  \bibinfo{journal}{Int. J. Bifurcation and Chaos Appl. Sci. Eng.}
  \textbf{\bibinfo{volume}{17}}, \bibinfo{pages}{3397} (\bibinfo{year}{2007}).

\bibitem[{\citenamefont{Kapitaniak and Wojewoda}(1993)}]{ufsk2006}
  \bibinfo{author}{\bibfnamefont{U.}~\bibnamefont{Feudel}},
\bibinfo{author}{\bibfnamefont{S.}~\bibnamefont{Kuznetsov}} \bibnamefont{and}
  \bibinfo{author}{\bibfnamefont{A.}~\bibnamefont{Pikovsky}},
  \emph{\bibinfo{title}{Strange nonchaotic attractors: Dynamics between order
  and chaos in quasiperiodically forced systems}}
  (\bibinfo{publisher}{World Scientific}, \bibinfo{address}{singapore},
  \bibinfo{year}{2006}).

\bibitem[{\citenamefont{Zhou et~al.}(1992)\citenamefont{Zhou, Moss, and
  Bulsara}}]{tzfm1992}
\bibinfo{author}{\bibfnamefont{T.}~\bibnamefont{Zhou}},
  \bibinfo{author}{\bibfnamefont{F.}~\bibnamefont{Moss}}, \bibnamefont{and}
  \bibinfo{author}{\bibfnamefont{A.}~\bibnamefont{Bulsara}},
  \bibinfo{journal}{Phys. Rev. A} \textbf{\bibinfo{volume}{45}},
  \bibinfo{pages}{5394} (\bibinfo{year}{1992}).

\bibitem[{\citenamefont{Ding et~al.}(1997)\citenamefont{Ding, Deutsch,
  Dingklage, and Wilke}}]{wxdhd1997}
\bibinfo{author}{\bibfnamefont{W.~X.} \bibnamefont{Ding}},
  \bibinfo{author}{\bibfnamefont{H.}~\bibnamefont{Deutsch}},
  \bibinfo{author}{\bibfnamefont{A.}~\bibnamefont{Dinklage}},
  \bibnamefont{and} \bibinfo{author}{\bibfnamefont{C.}~\bibnamefont{Wilke}},
  \bibinfo{journal}{Phys. Rev. E} \textbf{\bibinfo{volume}{55}},
  \bibinfo{pages}{3769} (\bibinfo{year}{1997}).

\bibitem[{\citenamefont{Ketoja and Satija}(1997)}]{jakis1997}
\bibinfo{author}{\bibfnamefont{J.~A.} \bibnamefont{Ketoja}} \bibnamefont{and}
  \bibinfo{author}{\bibfnamefont{I.}~\bibnamefont{Satija}},
  \bibinfo{journal}{Physica D} \textbf{\bibinfo{volume}{109}},
  \bibinfo{pages}{70} (\bibinfo{year}{1997}).

\bibitem[{\citenamefont{Ruiz and Parmananda}(2007)}]{ruiz2007}
\bibinfo{author}{\bibfnamefont{G.}~\bibnamefont{Ruiz}} \bibnamefont{and}
  \bibinfo{author}{\bibfnamefont{P.} \bibnamefont{Parmananda}},
  \bibinfo{journal}{Phys. Lett. A} \textbf{\bibinfo{volume}{367}},
  \bibinfo{pages}{478} (\bibinfo{year}{2007}).
 
\bibitem[{\citenamefont{Thamilmaran et~al.}(2006)\citenamefont{Thamilmaran,
  Senthilkumar, Venkatesan, and Lakshmanan}}]{sgps2004}
\bibinfo{author}{\bibfnamefont{S.}~\bibnamefont{Graziani}},
  \bibinfo{author}{\bibfnamefont{P.} \bibnamefont{Silar}},
  \bibnamefont{and} \bibinfo{author}{\bibfnamefont{M.~J}~\bibnamefont{Daboussi}},
  \bibinfo{journal}{BMC Biology}
  \textbf{\bibinfo{volume}{2}}, \bibinfo{pages}{18} (\bibinfo{year}{2004}).

\bibitem[{\citenamefont{Thamilmaran et~al.}(2006)\citenamefont{Thamilmaran,
  Senthilkumar, Venkatesan, and Lakshmanan}}]{ddrl2006}
\bibinfo{author}{\bibfnamefont{D.}~\bibnamefont{Dubnau}},
  \bibnamefont{and} \bibinfo{author}{\bibfnamefont{R}~\bibnamefont{Losick}},
  \bibinfo{journal}{Molecular Microbiology}
  \textbf{\bibinfo{volume}{61}}, \bibinfo{pages}{564} (\bibinfo{year}{2006}).  

\bibitem[{\citenamefont{Thamilmaran et~al.}(2006)\citenamefont{Thamilmaran,
  Senthilkumar, Venkatesan, and Lakshmanan}}]{arbej1991}
\bibinfo{author}{\bibfnamefont{A. R.}~\bibnamefont{Bulsara}},
\bibinfo{author}{\bibfnamefont{E.}~\bibnamefont{Jacobs}},
\bibinfo{author}{\bibfnamefont{T.}~\bibnamefont{Zhou}},
\bibinfo{author}{\bibfnamefont{F.}~\bibnamefont{Moss}},
  \bibnamefont{and} \bibinfo{author}{\bibfnamefont{L.}~\bibnamefont{Kiss}},
  \bibinfo{journal}{J. Theor. Biol.}
  \textbf{\bibinfo{volume}{154}}, \bibinfo{pages}{531} (\bibinfo{year}{1991});
\bibinfo{author}{\bibfnamefont{A.}~\bibnamefont{Longtin}},
\bibinfo{author}{\bibfnamefont{A.}~\bibnamefont{Bulsara}},
\bibinfo{author}{\bibfnamefont{D.}~\bibnamefont{Pierson}},
  \bibnamefont{and} \bibinfo{author}{\bibfnamefont{F.}~\bibnamefont{Moss}},
  \bibinfo{journal}{Biol. Cybern.}
  \textbf{\bibinfo{volume}{70}}, \bibinfo{pages}{569} (\bibinfo{year}{1994}).
  
\bibitem[{\citenamefont{Thamilmaran et~al.}(2006)\citenamefont{Thamilmaran,
  Senthilkumar, Venkatesan, and Lakshmanan}}]{vcmm2000}
\bibinfo{author}{\bibfnamefont{V.}~\bibnamefont{Chinarov}},
  \bibnamefont{and} \bibinfo{author}{\bibfnamefont{M.}~\bibnamefont{Menzinger}},
  \bibinfo{journal}{BioSystems}
  \textbf{\bibinfo{volume}{55}}, \bibinfo{pages}{137} (\bibinfo{year}{2000}).    

\bibitem[{\citenamefont{Thamilmaran et~al.}(2006)\citenamefont{Thamilmaran,
  Senthilkumar, Venkatesan, and Lakshmanan}}]{zmgwyl2004}
\bibinfo{author}{\bibfnamefont{Z. M.}~\bibnamefont{Ge}},
  \bibnamefont{and} \bibinfo{author}{\bibfnamefont{W. Y.}~\bibnamefont{Leu}},
  \bibinfo{journal}{Chaos, Solitons and Fractals}
  \textbf{\bibinfo{volume}{20}}, \bibinfo{pages}{502} (\bibinfo{year}{2004}). 
  
\bibitem[{\citenamefont{Thamilmaran et~al.}(2006)\citenamefont{Thamilmaran,
  Senthilkumar, Venkatesan, and Lakshmanan}}]{gdvrr1998}
\bibinfo{author}{\bibfnamefont{G. D.}~\bibnamefont{VanWiggeren}},
  \bibnamefont{and} \bibinfo{author}{\bibfnamefont{R}~\bibnamefont{Roy}},
  \bibinfo{journal}{Science}
  \textbf{\bibinfo{volume}{279}}, \bibinfo{pages}{1198} (\bibinfo{year}{1998});
  \bibinfo{journal}{Phys. Rev. Lett.}
  \textbf{\bibinfo{volume}{81}}, \bibinfo{pages}{3547} (\bibinfo{year}{1998}). 

\bibitem[{\citenamefont{Thamilmaran et~al.}(2005)\citenamefont{Thamilmaran,
  Senthilkumar, Lakshmanan, and Ahmed}}]{ktdvsk2005}
\bibinfo{author}{\bibfnamefont{K.}~\bibnamefont{Thamilmaran}},
  \bibinfo{author}{\bibfnamefont{D.~V.} \bibnamefont{Senthilkumar}},
  \bibinfo{author}{\bibfnamefont{M.}~\bibnamefont{Lakshmanan}},
  \bibnamefont{and} \bibinfo{author}{\bibfnamefont{A.}~\bibnamefont{Ishaq Ahmed}},
  \bibinfo{journal}{Int. J. Bifurcation and Chaos Appl. Sci. Eng.}
  \textbf{\bibinfo{volume}{15}}, \bibinfo{pages}{2} (\bibinfo{year}{2005}).


\end{thebibliography}
\end{document}